\documentstyle[psfig]{mn}

\begin{document}
\title[Baryon fractions of clusters]
{Constraints on the asymptotic baryon fractions of galaxy clusters
at large radii}
\author[Wu and Xue]{Xiang-Ping Wu$^{1,2}$ and Yan-Jie Xue$^{1,2}$\\
%\footnotesize{
$^1$Beijing Astronomical Observatory, Chinese Academy of Sciences,
    Beijing 100012, China\\
$^2$National Astronomical Observatories, Chinese Academy of Sciences,
    Beijing 100012, China\\
}
%\begin{document}
\date{submitted 1999 June 28; accepted 1999 September 13}
\maketitle
\begin{abstract}
While X-ray measurements have so far revealed an increase in 
the volume-averaged baryon fractions $f_b(r)$ of galaxy clusters
with cluster radii $r$,  $f_b(r)$ should asymptotically reach 
a universal value $f_b(\infty)=f_b$,
provided that clusters are representative of the Universe.
In the framework of hydrostatic equilibrium for intracluster gas, 
we have derived the necessary conditions for $f_b(\infty)=f_b$: 
The X-ray surface brightness profile described by the $\beta$ model
and the temperature profile approximated by the polytropic model should 
satisfy $\gamma\approx2(1-1/3\beta)$ and $\gamma\approx1+1/3\beta$ for 
$\beta<1$ and $\beta>1$, respectively, which sets a stringent 
limit to the polytropic index: $\gamma<4/3$. In particular, 
a mildly increasing temperature with radius is required if the 
observationally fitted $\beta$ parameter is in the range $1/3<\beta<2/3$.
It is likely that a reliable determination of the universal baryon fraction
can be achieved in the small $\beta$ clusters  because 
the disagreement between the exact and asymptotic 
baryon fractions for clusters with  $\beta>2/3$ breaks down 
at rather large radii ($\ga30r_c$) where
hydrostatic equilibrium has probably become inapplicable.  
We further explore how to obtain the asymptotic value $f_b(\infty)$
of baryon fraction  from the X-ray measurement made primarily 
over the finite central region of a cluster. We demonstrate our method  
using a sample of 19 strong lensing clusters, which enables us to  
place a useful constraint on $f_b(\infty)$:  
$0.094\pm0.035 \leq f_b(\infty) \leq 0.41\pm0.18$, corresponding to
a cosmological density parameter
$0.122\pm0.069 \leq \Omega_M \leq 0.53\pm0.28$ for 
$H_0=50$ km s$^{-1}$ Mpc$^{-1}$. 
An optimal estimate of $f_b(\infty)$ based on three cooling flow clusters
with $\beta<1/2$ in our lensing cluster sample
yields  $\langle f_b(\infty)\rangle = 0.142\pm0.007$ or
$\Omega_M = 0.35\pm0.09$. 
\end{abstract}


\vskip -3in

\section{Introduction}

In the standard scenario of structure formation, a typical galaxy cluster  
draws its matter (baryon + nonbaryon) from a region of radius of
$\sim20$ Mpc in the Universe. Therefore, it is widely believed that
clusters should be fair samples of baryonic and nonbaryonic matter 
compositions, and thus their baryon fractions 
$f_b$ can be used to determine the average mass density of the Universe, 
$\Omega_M$, in conjunction with the Big Bang Nucleosynthesis (BBN)
(e.g. White et al. 1993; David, Jones \& Forman 1995).
However, all the X-ray measurements have so far shown an increase in
the baryon fractions $f_b(r)$ of clusters with radii and no any evidence for
an asymptotic tendency towards a universal value at large radii
(White \& Fabian 1995; Ettori, Fabian \& White 1997; David 1997; 
White, Jones \& Forman 1997; Ettori \& Fabian 1999; 
for a recent summary see Wu 1999a). 
The conflict between X-ray measurement and theoretical expectation
becomes even more serious when the observed temperature profiles 
$T(r)$, which often exhibit a radial decline at large radii and can be
well approximated by the polytropic models of $\gamma\approx1.1$--$1.3$
(Markevitch et al. 1998; Ettori \& Fabian 1999), 
instead of an isothermal gas distribution are used (Henriksen \& Mamon
1994; Henriksen \& White 1996; Markevitch et al. 1999). 
In particular, it has been realized that 
the puzzle is unlikely to be associated with the conventional 
cluster mass estimates at least within the Abell radius,  
which relies upon the hydrostatic equilibrium hypothesis 
for the dynamical state of clusters in the computation of
their total masses. This point has been recently justified by the excellent
agreement among the X-ray, optical and weak lensing determined 
cluster masses on scales $0.5\la r \la 3$ Mpc
(Allen 1998; Wu et al. 1998), where and also 
hereafter the Hubble constant is taken to be 
$H_0=50$ km s$^{-1}$ Mpc$^{-1}$. 
Yet, we cannot exclude the possibility that the regions 
accessible to current observations are still not large enough for 
the volume-averaged $f_b(r)$ in clusters to be representative of the
Universe. Meanwhile, it remains unclear  how accurate  
the conventional cluster mass estimates will be on 
the outskirts of clusters where the infalling matter probably becomes 
important. A definite resolution to the puzzle may require
detailed studies of these external regions.
A good example of such studies has been  
provided in optical by Geller, Diaferio \& Kurtz (1999) for the Coma cluster
although their work yields no information about the baryonic mass of 
the cluster.

In this paper we make no attempt at resolving the puzzle. Instead,
we first study the necessary conditions for $f_b(\infty)=f_b$ if
clusters are the well dynamically-relaxed systems and  
share a common value of baryon fraction $f_b$ 
at $r \rightarrow \infty$. We then explore the possibility of deriving
the asymptotic value $f_b(\infty)$  from a local measurement 
$f_b(r)$ made primarily over the central region of a cluster.
Finally, we apply our method to a strong lensing cluster sample and 
demonstrate how a useful constraint can be set on the universal 
value $f_b$.

\section{BARYON FRACTION}

\subsection{Volume-averaged baryon fraction}

Since the contribution of stellar mass $M_{*}$ to the baryon fraction 
of a cluster is typically 5 times smaller than that of gas mass $M_{gas}$ 
(e.g. White et al. 1993; Ettori et al. 1997), 
we will not include $M_{*}$
in the estimate of the cluster baryon fraction below. 
Following the conventional treatment
(Cowie, Henriksen \& Mushotzky 1987; 
Henriksen \& Mamon 1994), we assume a $\beta$ model for 
intracluster gas characterized by
the electron number density profile, $n_e(r)=n_{e0}[1+(r/r_c)^2]^{-\delta}$,
and an equation of state, $T(r)=T_0[n_e(r)/n_{e0}]^{\gamma-1}$, where
$n_{e0}$ and $T_0$ are the corresponding central values and 
$\delta\approx3\beta/2$. The total mass in gas within a sphere of radius $r$
is simply 
%1,2
\begin{eqnarray}
M_{gas}(x)  = & 4\pi \mu_e m_p n_{e0}r_c^3 \tilde{M}_{gas}(x),\\
\tilde{M}_{gas}(x) = &  \int_0^{x}\frac{y^2dy}{(1+y^2)^{\delta}},
\end{eqnarray}
where $x=r/r_c$, $\mu_e=2/(1+X)$, and $X=0.768$ is 
the hydrogen mass fraction in the
primordial abundances of hydrogen and helium. 
If the X-ray emitting gas is in
hydrostatic equilibrium with the underlying gravitational potential of the
cluster, then the total dynamical mass within $r$ (or $x$) is
%3
\begin{eqnarray}
M_{tot}(r) & =-\frac{kTr}{G\mu_im_p}
               \left(\frac{d\ln n_e}{d\ln r}+\frac{d\ln T}{d\ln r}\right)
			\nonumber  \\
           & = 2\gamma\delta\frac{kT_0r_c}{G \mu_i m_p} 
               \frac{x^3}{(1+x^2)^{1+\delta(\gamma-1)}},
\end{eqnarray}
in which $\mu_i=0.585$ denotes the mean molecular weight. The volume-averaged
baryon fraction within $x$ is then 
%4
\begin{equation}
f_b(x)= \frac{M_{gas}(x)}{M_{tot}(x)}=
f_0(1+x^2)^{1+\delta(\gamma-1)}\frac{\tilde{M}_{gas}(x)}{x^3},
\end{equation}
in which
%5
\begin{equation}
f_0   =  \frac{2\pi G \mu_e\mu_i m_p^2 n_{e0} r_c^2}{\gamma\delta kT_0}.
\end{equation}

We now examine the dependence of baryon fraction on cluster radius. In the
central region where $x\rightarrow 0$, 
%6
\begin{equation}
f_b(0)=\frac{f_0}{3}.
\end{equation}
At large radius where $x\gg1$ the baryon fraction varies asymptotically as
%7,8
\begin{eqnarray}
f_b(x)\approx f_0C x^{2\delta(\gamma-1)-1}+
               \frac{f_0}{3-2\delta}x^{2-2\delta+2\delta(\gamma-1)}, & 
                      \delta \neq \frac{3}{2};\\
f_b(x)\approx f_0(\ln 2x -1)x^{3(\gamma-1)-1},  & \delta=\frac{3}{2},
\end{eqnarray}
where $C$ is a constant. This disagrees with the previous result given by
Henriksen \& Mamon (1994) who claimed 
$f_b(x)\sim x^{2-\delta[2-(\gamma-1)]}$ at large radius, which results in
$\gamma=3-2/\delta$ if $f_b(\infty)$ is asymptotically constant. 
It is likely that they have mistaken  $x^{2+\delta(\gamma-1)}$ for the
asymptotic expansion of $(1+x^2)^{1+\delta(\gamma-1)}$, and oversimplified
the integral $\tilde{M}_{gas}(x)$. For instance, their analysis has neglected
the asymptotic behavior of $\tilde{M}_{gas}(x)\propto\ln x$ at large radius 
for $\delta=3/2$. Indeed, in the single $\beta$ model
with $\delta=3/2$ or $\beta\approx1$, $f_b(\infty)$ appears to be divergent.
For any other $\delta$ (i.e. $\delta\neq3/2$), 
if we require $f_b(\infty)$ to be 
constant, it is easy to show that the polytropic index 
$\gamma$ should satisfy 
%9
\begin{equation}
\gamma=\left\{
\begin{array}{ll}
2-\frac{1}{\delta}, & \delta<\frac{3}{2};\\
1+\frac{1}{2\delta}, & \delta>\frac{3}{2}.
\end{array} \right.
\end{equation}
This sets a stringent limit to the value of $\gamma$ 
%10
\begin{equation}
\gamma<\frac{4}{3}.
\end{equation}
Namely, the necessary condition for 
the baryon fraction of a cluster asymptotically approaching a
universal value at large radius is that the intracluster gas has 
a polytropic index $\gamma<4/3$. It appears that all the observed
temperature profiles of clusters so far have indeed met  
this simple requirement (e.g. Markevitch et al. 1998). Furthermore,  because
in the great majority of cases  the observed X-ray surface brightness 
profiles show $\beta<1$,  we will only focus on the situation of 
where $\delta\approx3\beta/2<3/2$,  for which the relationship between 
the volume-averaged baryon fractions measured at $r=0$ and $r=\infty$, 
according to eqs.(6)-(9), is 
%11
\begin{equation}
f_b(\infty)=\frac{3}{3-2\delta}f_b(0)\approx\frac{1}{1-\beta}f_b(0).
\end{equation}
Nevertheless, the necessary conditions eq.(9) will yield
an unphysical value of $\gamma\leq0$ if $\delta\leq1/2$, or an unusual
result of $0<\gamma<1$ if $1/2<\delta<1$. The former may be avoided
because the observationally fitted $\beta$ parameters from 
the X-ray surface brightness profiles of clusters are usually larger 
than $1/3$, i.e., $\delta\ga 1/2$. In the latter case ($1/3<\beta<2/3$), 
the temperature profile of a cluster is required to slightly increase with 
outward radius in order for $f_b(\infty)$ to maintain constant.
This last point is indeed beyond our natural expectations.
Whether or not such a requirement is consistent with the 
spectroscopic data will be addressed in the discussion section.

It deserves to examine the issue as to how fast the baryon 
fraction $f_b(x)$ approaches the universal value. A relevant question
is:  to what radius can we take the volume-averaged baryon fraction $f_b(x)$ 
as a good approximation of the universal value ?  We display in Fig.1 
the variation of $f_b(x)$ with radius for $\delta=3/4$, $1$ and
$1.275$, respectively, along with our asymptotic approximations 
from eq.(11).
Surprisingly, the disagreement between the exact and asymptotic 
baryon fractions breaks down at rather larger radii, especially
for clusters with large $\delta$. Consequently, it is unlikely 
that one can `directly' measure the cosmic baryon fraction 
within a cluster of $\delta>1$ (or $\beta>2/3$), 
which needs the detection of X-ray emission out to a radius of $r\ga 30r_c$,
corresponding to $r\ga 7.5$ Mpc for an average
X-ray core radius of $0.25$ Mpc. 
This even does not account for the fact that hydrostatic equilibrium
fails in this external region. Indeed, the problem becomes much more
serious for larger $\beta$ clusters because 
the observationally fitted $\beta$ parameter is strongly correlated 
with $r_c$, with a large value of $\beta$ giving rise to a large $r_c$ value. 
Moreover, it is apparent from Fig.1 that $f_b(x)$ increases 
monotonically with radius within the region accessible to current 
observations ($r/r_c\la10$), which may provide a reasonable explanation 
for the present status of X-ray measurements of cluster baryon fractions 
mentioned at the very onset. We note, however, that 
it should be possible to estimate the universal baryon fraction from 
the small $\beta$ clusters. For example,  in a $\beta=1/2$ cluster 
the deviations of the baryon fractions measured within $r=10r_c$ and $5r_c$ 
from the asymptotic value  
are only $5.5\%$ and $11.5\%$, respectively. It will be expected that
a consistent baryon fraction can be obtained and applied to the determination
of the cosmic density parameter $\Omega_M$ when an ensemble of small $\beta$
clusters are used.

\subsection{Projected baryon fraction}

Next we compute the projected cluster baryon fraction 
$\tilde{f}_b(b)=m_{gas}(b)/m_{tot}(b)$ within radius $b$ along 
the line of sight, where $m_{gas}(b)$ and $m_{tot}(b)$ are 
the projected gas and gravitating mass of 
the cluster within $b$, respectively. Under the same working hypothesis
as in the above discussion, we have
%12
\begin{equation}
m_{gas}(x)=  4\pi\mu_em_p n_{e0}r_c^3 \tilde{m}_{gas}(x),
\end{equation}
\begin{eqnarray*}
\tilde{m}_{gas}(x)= &
\tilde{M}_{gas}(x)+\int_x^{\infty}(y-\sqrt{y^2-x^2})
      \frac{y}{(1+y^2)^{\delta}}dy, 
\end{eqnarray*}
and
%13
\begin{equation}
m_{tot}(x)=  2\gamma\delta\frac{kT_0r_c}{G\mu_im_p} \tilde{m}_{tot}(x),
\end{equation}
\begin{eqnarray*}
\tilde{m}_{tot}(x)= &
           \frac{x^3}{(1+x^2)^{1+\delta(\gamma-1)}} +   \\
            &  \int_x^{\infty}
             \left[3-[1+\delta(\gamma-1)]\frac{2y^2}{1+y^2}\right]
             \frac{y(y-\sqrt{y^2-x^2})}{(1+y^2)^{1+\delta(\gamma-1)}} dy, 
\end{eqnarray*}
where $x=b/r_c$. The projected baryon fraction within $x$ is then
%14
\begin{equation}
\tilde{f}_b(x)=\frac{m_{gas}(x)}{m_{tot}(x)}=
f_0\frac{\tilde{m}_{gas}(x)}{\tilde{m}_{tot}(x)}.
\end{equation}
As $x$ approaches $0$, we have
\begin{equation}
%15
\tilde{f}_b(0)= f_0
             \frac{\int_0^{\infty}\frac{dy}{(1+y^2)^{\delta}}}
                  {\int_0^{\infty}
             \{3-[(1+\delta(\gamma-1)]\frac{2y^2}{1+y^2}\}
             \frac{dy}{(1+y^2)^{1+\delta(\gamma-1)}} }. 
\end{equation}
In particular, if we take the condition, $\gamma=2-1/\delta$, found in
the above subsection for $\delta<3/2$ in order for $f_b(\infty)$ 
to become a universal value, it can be shown that 
the above expression in the case of $\delta\geq1/2$ (i.e. $\beta\ga 1/3$)
reduces to
%16
\begin{equation}
\tilde{f}_b(0)=\frac{f_0}{2}.
\end{equation}
As a result,  in the limits of $1/3<\beta<1$ and 
$\gamma<4/3$, we find according to eqs.(6), (11) and (16) 
%17
\begin{equation}
f_b(\infty)=\frac{2}{3-2\delta}\tilde{f}_b(0)
\approx\frac{2}{3}\frac{\tilde{f}_b(0)}{1-\beta}.
\end{equation}

\section{APPLICATION TO STRONG LENSING CLUSTERS}

In this section we demonstrate how to constrain the asymptotic baryon 
fraction of a cluster at large radius using the X-ray measurement made
within a limiting detection radius. For this purpose, 
we select clusters from the Strong Lensing Cluster Sample 
compiled by Wu et al (1998), which contains 48 arclike images of
background galaxies gravitationally lensed by 38 foreground clusters.
Here we exclude the clusters (1)whose X-ray temperatures are unknown or 
given by indirect methods such as the $L_x$-$T$ and $\sigma$-$T$ 
correlations, where $L_x$ is the X-ray luminosity and 
$\sigma$ is the velocity dispersion of 
cluster galaxies, or (2)whose X-ray surface brightness profiles are not
well fitted by the $\beta$ models due to either the poor data quality 
(e.g. Cl2244, etc.) or other mechanisms intrinsic to clusters. 
Nevertheless, we have included the well-known arc-cluster A370 but assumed 
a core radius $r_c=0.25$ Mpc and $\beta=2/3$ although its X-ray surface
brightness is not well constrained (e.g. Ota, Mitsuda \& Fukazawa 1998; 
Arnaud \& Evrard 1999). 
This results in a sample of 19 clusters, in which there are 25 
strongly distorted images of distant galaxies (Table 1). 
The projected cluster mass interior to the position ($r_{arc}$) 
of an arclike image can be simply estimated through
%18
\begin{equation}
m_{lens}(r_{arc})=\pi r_{arc}^2 \Sigma_{crit},
\end{equation}
where $\Sigma_{crit}=(c^2/4\pi G)(D_s/D_dD_{ds})$ is the critical surface
mass density, with $D_d$, $D_s$ and $D_{ds}$ being the angular diameter
distances to the cluster, to the background galaxy, and from the
cluster to the galaxy, respectively. Spectroscopic data have not been
available for about half of the arcs, for which we assume a redshift of
$z_s=0.8$.  The uncertainties of $m_{lens}$ due to the unknown 
redshifts of arcs are very minor. While the strong lensing can indeed 
yield the projected gravitating mass of a cluster independently of 
cluster matter contents and their dynamical state, eq.(18) may 
lead to an overestimate of the true cluster mass 
by a factor of $\sim2$--$4$ because it does not take 
the contribution of substructures into account.  
The factor of $\sim2$--$4$ comes 
from a statistical comparison of cluster masses determined from various
methods including strong lensing, weak lensing, X-ray, optical and 
numerical simulations (Allen 1998; Wu et al. 1998; Wu
1999b).  Consequently, the ratio of the projected mass in gas 
$m_{gas}(r_{arc})$ to the strong lensing derived mass $m_{lens}(r_{arc})$
provides a low limit to the true baryon fraction 
$f_b(r_{arc})$ within $r_{arc}$.
Moreover,  since all the arclike images are detected within the central 
regions of the clusters, we can use $\tilde{f}_b(r_{arc})$ 
to approximately represent the projected baryon fractions 
at cluster centers, $\tilde{f}_b(0)\approx\tilde{f}_b(r_{arc})$. That is, 
%19
\begin{equation}
\tilde{f}_b(0)\approx \tilde{f}_b(r_{arc}) 
\geq \frac{m_{gas}(r_{arc})}{m_{lens}(r_{arc})}.
\end{equation} 
The resultant values of $m_{gas}(r_{arc})$, $m_{lens}(r_{arc})$ and 
their ratios within the positions of 25 arcs 
have been given in Table 1 for a flat cosmological model $\Omega_0=1$.

\begin{table*}
\vskip 0.2truein
\begin{center}
\caption{Strong Lensing Cluster Sample$^a$}
\vskip 0.2truein
%\begin{scriptsize}
\begin{tabular}{lclllcllclcl}
\hline
cluster  & $z_{cluster}$ & $T$ (keV) & 
$L_x^b$ & $r_c^c$ & $\beta$ & $z_{arc}$ & 
$r_{arc}^c$   &  $m_{gas}^d$ &  $m_{lens}^d$  & 
$m_{gas}/m_{lens}(\%)$ & $f_0(\%)^e$ \\
\hline
A370 & 0.373 & $7.13^{+1.05}_{-1.05}$ & $20.8^{+2.0}_{-2.0}$ & 0.25 & 0.67  &  
     1.3 & 0.35  & $2.81^{+0.25}_{-0.23}$  &  130 & $2.16^{+0.19}_{-0.18}$ &
                                                    $18.6^{+5.2}_{-3.7}$ \\
     &       &                     &                      &      &      &   
     0.724 & 0.16 &$0.73^{+0.07}_{-0.06}$ &  29.0 & $2.52^{+0.23}_{-0.21}$ & \\
A963 & 0.206 & $6.13^{+0.45}_{-0.30}$ & $16.1^{+1.8}_{-1.8}$ & 0.08 & 0.67  &
     ... &0.0517 & $0.12^{+0.01}_{-0.01}$ &  2.5 & $4.91^{+0.33}_{-0.36}$ &
                                                    $11.1^{+1.4}_{-1.5}$  \\
     &        &                      &                    &      &      & 
     0.711 &0.080 &$0.27^{+0.02}_{-0.02}$ &  6.0 & $4.44^{+0.30}_{-0.33}$ & \\
A1689 & 0.181 & $9.02^{+0.40}_{-0.30}$ & $50.5^{+1.1}_{-1.1}$& 0.13 & 0.65 & 
     ... & 0.183 & $1.56^{+0.03}_{-0.03}$ & 36  & $4.32^{+0.08}_{-0.09}$ &
                                                   $15.0^{+0.8}_{-0.9}$   \\
A1835 & 0.252 & $9.80^{+1.40}_{-1.40}$ & $105^{+14}_{-14}$ & 0.074 & 0.65 &
     ... & 0.150 & $1.65^{+0.18}_{-0.17}$ &19.8 & $8.35^{+0.89}_{-0.83}$ &
                                                   $14.8^{+4.3}_{-3.1}$   \\
A2163 & 0.203 & $14.60^{+0.85}_{-0.85}$ & $133^{+19}_{-19}$ & 0.34 & 0.64 &
   0.728 &0.0661 & $0.24^{+0.02}_{-0.02}$ & 4.1  & $5.84^{+0.50}_{-0.51}$ &
                                                   $21.4^{+3.3}_{-2.9}$  \\
A2218 & 0.171 & $7.10^{+0.20}_{-0.20}$&$18.1^{+0.75}_{-0.75}$&0.23 &0.65 &  
   1.034 & 0.26  & $1.61^{+0.04}_{-0.04}$ & 27 & $5.98^{+0.17}_{-0.17}$ &
                                                   $16.1^{+0.9}_{-0.9}$  \\
      &       &                    &                      &      &      &  
   0.702 &0.0794 & $0.18^{+0.01}_{-0.01}$ & 6.23 & $2.92^{+0.08}_{-0.08}$ &  \\
      &       &                    &                      &      &      &  
   0.515 & 0.0848& $0.21^{+0.01}_{-0.01}$ & 5.70 & $3.63^{+0.10}_{-0.10}$ &  \\
A2219 & 0.228 & $12.40^{+0.50}_{-0.50}$ & $64.6^{+7.0}_{-7.0}$ & 0.15& 0.64  &
    ...  & 0.079 & $0.35^{+0.02}_{-0.02}$ & 5.6  & $6.82^{+0.43}_{-0.44}$ &
                                                    $12.1^{+1.3}_{-1.2}$  \\
      &       &                      &                    &      &      & 
    ...  & 0.110 & $0.66^{+0.04}_{-0.04}$ & 16.0 & $4.41^{+0.26}_{-0.27}$ &  \\
A2390 & 0.228 & $11.10^{+1.00}_{-1.00}$ & $63.5^{+14.9}_{-14.9}$ & 0.22&0.66 & 
    0.913& 0.177 & $1.43^{+0.20}_{-0.21}$ & 25.4 & $5.64^{+0.78}_{-0.81}$ &
                                                    $17.3^{+4.3}_{-3.7}$  \\
A2744 & 0.308  & $11.00^{+0.50}_{-0.50}$  & $64.4^{+14.4}_{-14.4}$ &0.45&0.65 &
    ... & 0.1196 & $0.51^{+0.06}_{-0.07}$ & 11.36& $4.48^{+0.53}_{-0.58}$ &
                                                    $24.8^{+4.3}_{-4.1}$  \\
AC114  & 0.310 & $9.76^{+1.04}_{-0.85}$&$38.1$               & 0.30 & 0.67  &
   1.86 & 0.0676 & $3.47^{+0.08}_{-0.09}$  &  2.98 &$2.67^{+0.06}_{-0.07}$ &
                                                    $18.7^{+2.3}_{-2.1}$ \\
       &       &                    &                     &      &      & 
   0.639& 0.35   & $0.16^{+0.00}_{-0.00}$  & 130   &$5.42^{+0.12}_{-0.14}$ & \\
Cl0500 & 0.327 & $7.20^{+3.70}_{-1.80}$ & $17.5^{+1.3}_{-0.9}$ & 0.030 &0.36 & 
    ... & 0.15 & $1.21^{+0.14}_{-0.15}$   & 19.0   & $6.37^{+0.07}_{-0.08}$ &
                                                    $1.3^{+0.6}_{-0.5}$ \\
MS0440 & 0.197 & $5.30^{+1.27}_{-0.85}$ &$7.4^{+1.0}_{-1.0}$& 0.054& 0.68 &   
   0.530 &0.089& $0.23^{+0.03}_{-0.03}$  &  8.9  &   $2.60^{+0.29}_{-0.31}$ &
                                                    $7.5^{+2.5}_{-2.2}$ \\
MS0451 & 0.539 & $10.17^{+1.55}_{-1.26}$ & $53.7$            &  0.088 & 0.84 & 
     ... & 0.190 & $1.51^{+0.05}_{-0.05}$  & 52 &  $2.91^{+0.10}_{-0.10}$ &
                                                    $14.0^{+2.5}_{-2.3}$ \\
MS1008 & 0.306 & $7.29^{+2.45}_{-1.52}$& $9.1^{+1.2}_{-1.2}$ & 0.12 &0.68 &   
     ... & 0.26  & $1.18^{+0.15}_{-0.16}$  & 61 & $1.94^{+0.25}_{-0.16}$ &
                                                    $8.6^{+3.6}_{-3.0}$  \\
MS1358 & 0.329 & $6.5^{+0.7}_{-0.6}$ & $21.8^{+3.8}_{-3.8}$ & 0.023 & 0.36 &
    4.92 & 0.121 & $1.02^{+0.11}_{-0.12}$  & 8.27 & $12.4^{+1.3}_{-1.4}$ &
                                                    $1.4^{+0.3}_{-0.3}$  \\
MS1455 & 0.257 & $5.45^{+0.29}_{-0.28}$ & $29.4^{+1.5}_{-1.5}$ &0.040 &0.66 &  
     ... & 0.098 &$0.52^{+0.02}_{-0.02}$   & 8.6  & $6.04^{+0.24}_{-0.23}$&
                                                    $12.0^{+1.2}_{-1.0}$ \\
MS2137 & 0.313 & $4.37^{+0.38}_{-0.72}$ & $26.3^{+3.6}_{-3.6}$ & 0.054 & 0.64 &
     ... & 0.0874&$0.44^{+0.05}_{-0.04}$ &7.1  & $6.17^{+0.71}_{-0.56}$ &
                                                    $16.6^{+5.6}_{-2.7}$ \\
PKS0745 & 0.103 & $8.7^{+1.6}_{-1.2}$ &$57.2^{+4.4}_{-4.4}$&0.056& 0.59  &
   0.433 & 0.0459& $0.17^{+0.01}_{-0.01}$  & 3.0  & $5.81^{+0.45}_{-0.46}$ &
                                                    $9.4^{+2.3}_{-2.1}$ \\
RXJ1347 & 0.451 & $11.37^{+1.10}_{-0.92}$ &$198^{+22}_{-22}$& 0.024 & 0.42 &
    ... & 0.24  & $6.15^{+0.47}_{-0.48}$  & 42   &  $14.6^{+1.1}_{-1.2}$ &
                                                    $3.4^{+0.6}_{-0.6}$  \\
\hline
\end{tabular}
%\end{scriptsize}
\end{center}
\parbox{6.5in}{$^a$Lensing and X-ray data (except for $r_c$ and $\beta$) 
               are taken from Wu et al. (1998) 
               and Wu, Xue \& Fang (1999), respectively;} 
\parbox{6.5in}{$^b$(X-ray) bolometric luminosity in units 
               of $10^{44}$ erg s$^{-1}$;}\\
\parbox{6.5in}{$^c$in units of Mpc;}\\
\parbox{6.5in}{$^d$projected mass within $r_{arc}$ 
               in units of $10^{13}M_{\odot}$;}
\parbox{6.5in}{$^e$calculated by eq.(5).}
 \end{table*}

On the other hand, we can also calculate the volume-averaged baryon fraction
around $r=0$ for each lensing cluster using eqs.(5) and (6), 
along with the restriction $\gamma=2-1/\delta$. Unfortunately, 
the temperature profiles have not been obtained for these 
lensing clusters. Instead, present X-ray observations have only provided 
the global emission weighted-temperatures 
%20
\begin{equation}
\overline{T}=\frac{\int \alpha(T)T(r)n^2_e(r)r^2dr}
                  {\int \alpha(T)    n^2_e(r)r^2dr} ,
\end{equation}
where $\alpha(T)$ is the cooling function. When $\overline{T}$ is used 
to calculate $f_0$,  
we will either overestimate (for $\beta>2/3$) or underestimate
(for $\beta<2/3$) the values of $f_0$ 
because  $\overline{T}$ are smaller (greater) 
than the central values $T_0$
in polytropic models with $\beta>2/3$ ($\beta<2/3$). 
We have attempted to assign the upper and low limits to 
the central baryon fractions, respectively, for the 6 and 13 clusters 
with $\beta>2/3$ and $\beta<2/3$ in Table 1 according to
%21
\begin{equation}
f_b(0)=\frac{1}{3}f_0|_{T_0=\overline{T}}. 
\end{equation}
However, it appears that we cannot make a clear distinction between 
the resultant upper and low limits among the 19 clusters 
(see also Fig.2). This probably implies that the evaluation of 
$f_b(0)$ is little affected by the approximation of 
the emission weighted temperature to the central temperature.

The low limits $\tilde{f}_b(0)$ [eq.(19)] and the  
central baryon fractions $f_b(0)$ approximated by eq.(21) can be converted 
into the total baryon fractions of clusters $f_b(\infty)$
in terms of the relations established in the above section 
[eqs.(11) and (17)], namely,
%22
\begin{equation}
f_{b,lens} \leq f_b(\infty)\approx  f_{b,xray},
\end{equation}
where 
%23,24
\begin{eqnarray}
f_{b,lens}\equiv
\frac{2}{3(1-\beta)} \frac{m_{gas}(r_{arc})}{m_{lens}(r_{arc})},\\
f_{b,xray}\equiv 
\frac{1}{3(1-\beta)} f_0|_{T_0=\overline{T}}.
\end{eqnarray}
In Fig.2 we plot the results of $f_{b,lens}$ and $f_{b,xray}$ 
obtained from the 25 arclike images 
among 19 lensing clusters listed in Table 1.
Note that $f_{b,xray}$ is independent of arc position $r_{arc}$, 
and here we utilize $r_{arc}$ to represent the lensing cluster only.
It is apparent that the lensing and X-ray results are clearly
separated except for three clusters, Cl0500, MS1358 and RXJ1347.
Dispersion in $f_{b,lens}$ is relatively small, and 
averaging over the 25 data points yields
%25
\begin{equation}
f_b(\infty) \geq \langle f_{b,lens}\rangle = 0.094\pm0.035, 
\end{equation}
where the error bar has not accounted for the uncertainties of 
$\beta$ and $r_c$ arising from the $\beta$ model fitting.
The average value  $\langle f_{b,lens}\rangle$ increases only slightly 
if the cooling flow clusters alone are considered:
%26
\begin{equation}
\langle f_{b,lens}\rangle = 0.106\pm0.033. 
\end{equation}
The reason why we emphasize this result 
is that an excellent agreement between cluster masses determined from
strong lensing and X-ray measurements for cooling flow clusters
has been claimed by Allen (1998). If this is true,  we would expect 
$f_{b,lens}=f_b(\infty)$. Nevertheless, we will still adopt
eq.(25) for caution's sake: First, the recent study of Lewis et al. (1999)
on the cluster mass estimates using the CNOC cluster sample 
does not reveal the large discrepancy
between cooling and non-cooling flow clusters; Second, 
the values of $f_{b,lens}$ we have found for cooling and non-cooling
flow clusters are essentially consistent with each other (see Fig.2).

A glimpse of Fig.2 reveals that the three clusters
(Cl0500, MS1358 and RXJ1347) which give smaller 
values of $f_{b,xray}$ all have their $\beta$ values less than $1/2$.
This motivates us to examine the dependence of $f_{b,xray}$ upon $\beta$ 
for our cluster sample (Fig.3).  Indeed,  
the values of $f_{b,xray}$ derived from the volume-averaged
baryon fractions among clusters with $\beta>2/3$  
are systematically high. As has been pointed out in section 2.1, 
these asymptotic values cannot be used as
a reliable indicator of the universal baryon fraction  because 
the agreement between the exact and asymptotic values of the baryon fractions
occurs at rather large radii $r\ga 30r_c$ where the hydrostatic 
equilibrium hypothesis may have already broken down.  On the other hand,  
there are good reasons that the asymptotic values $f_{b,xray}$ 
obtained among Cl0500, MS1358 and RXJ1347
can be considered as a good approximation of the universal baryon 
fraction: First, the $\beta$ parameters for these three clusters 
are smaller than $1/2$,
which guarantees the condition $f_b(x)\approx f_{b,xray}$ within $\sim10r_c$;
Second, all these three clusters are classified as the cooling flow
clusters, for which the total cluster masses $M_{tot}$
could be determined relatively accurately from eq.(3) (Allen 1998);
Third, the asymptotic values $f_{b,xray}$ from the volume-averaged
baryon fractions among these three clusters  
are essentially consistent with the asymptotic values $f_{b,lens}$, and
%27,28
\begin{eqnarray}
\langle f_{b,lens}\rangle = 0.121\pm0.088;\\
\langle f_{b,xray}\rangle = 0.142\pm0.007.
\end{eqnarray}

As a conservative estimate, we utilize the average values 
$\langle f_{b,lens}\rangle$ and $\langle f_{b,xray}\rangle$ among
all the clusters to be the low and upper limits to the 
universal baryon fraction
%29
\begin{equation}
0.094\pm0.035 \leq f_b(\infty) \leq 0.41\pm0.18, 
\end{equation}
which, in combination with the BBN prediction (Walker et al. 1991),
corresponds to the following constraints on the cosmological 
density parameter 
%30
\begin{equation}
0.122\pm0.069 \leq \Omega_M \leq 0.53\pm0.28,  
\end{equation}
where the uncertainty of the BBN prediction, $\Delta\Omega_b=0.01$,
has been included,  and $\Omega_b$ is the average baryon mass density
of the Universe in units of the critical density for closure.  
If the average value $\langle f_{b,xray}\rangle$ in the three
cooling clusters with $\beta<1/2$ is adopted, we have
%31
\begin{equation}
\Omega_M = 0.35\pm0.09.
\end{equation}

Finally, we investigate whether the asymptotic baryon fractions (or limits)
$f_{b,lens}$ and $f_{b,xray}$ derived from the strong lensing clusters 
depend on X-ray temperature. This is important because 
variation of baryon fractions among clusters
of different temperatures is not allowed in the standard models of
structure formation.  For the time being we can only take 
a less serious approach to the problem in the sense that the exact
baryon fractions  of these lensing clusters have remained unknown. 
Fig.4 illustrates the dependence of $f_{b,lens}$ and  $f_{b,xray}$
upon the emission-weighted temperature for the 19 clusters in Table 1. 
The best-fitted $f_{b,lens}-\overline{T}$ 
and  $f_{b,xray}-\overline{T}$ relations are, respectively,
%32,33
\begin{eqnarray}
f_{b,lens}=10^{-1.38\pm0.26}\overline{T}^{0.39\pm0.28};\\
f_{b,xray}=10^{-0.87\pm0.35}\overline{T}^{0.47\pm0.38},
\end{eqnarray}
where $\overline{T}$ is in units of keV.
These relations are marginally consistent with 
no increase of baryon fractions with $\overline{T}$
as predicted in the standard models 
although they cannot be used as a direct evidence.
Similar result has recently been reported by Mohr, Mathiesen \& Evrard (1999).

\section{Discussion and conclusions}

The baryon fraction $f_b(r)$ of a cluster should be averaged over 
a sufficiently large cluster volume 
in order for $f_b(r)$ to be representative of the universal
value $f_b$. However, this is limited by the sensitivity of 
current detectors which can only probe a finite region of the X-ray cluster. 
We have thus studied the condition for $f_b=f_b(\infty)$ and also the 
possibility of deriving $f_b(\infty)$ from the central baryon fraction 
$f_b(0)$ or $\tilde{f}_b(0)$. It has been shown that the polytropic index of
intracluster gas should satisfy $\gamma<\frac{4}{3}$ 
if the baryon fraction of a cluster is required to asymptotically reach 
a universal value $f_b$ at large radius. It appears that 
all the clusters whose temperature profiles have been available thus far   
indeed meet this simple requirement. 
We have found that clusters with small $\beta$ parameters
are likely to provide a better estimate of the universal baryon fraction,
which arises because the exact and asymptotic forms 
of the baryon fractions merge at rather large radii $r\ga 30r_c$ 
for $\beta\ga 2/3$, where hydrostatic equilibrium becomes to be
questionable.  
We have also demonstrated how to set useful constraints on
the baryon fraction $f_b$, and thereby the average mass density of 
the Universe, $\Omega_M$, using a sample of 19 strong lensing clusters. 
Our analysis gives support to a medium  density universe, and 
the scatters in the present estimate of $\Omega_M$ can be greatly reduced 
when a large cluster sample is employed.

However, there are several factors which should be taken seriously.
Our major concern is the current working model for clusters,
in which the hydrostatic equilibrium hypothesis and the conventional 
$\beta$ model for intracluster gas are 
extrapolated to sufficiently large radii. 
The availability of this working model to the Abell radius has been 
demonstrated by the reasonable agreement between different 
mass estimators.  In principle,  
clusters can approximately be regarded as dynamically-relaxed 
systems out to the falling shock radii where cluster matter is mixed 
with the infalling matter from background universe. 
Presumably, our conclusion cannot be applied beyond these radii, 
within which the volume-averaged matter composition should be of 
cosmological significance. Even so, since these boundaries are well beyond
the regions accessible to present observations,  it is unclear 
whether the intracluster gas in the outmost regions still follows 
the $\beta$ and polytropic models.

To ensure the universal constancy of the cluster baryon fraction at
large radius,  the intracluster gas should have a polytropic index 
of $0<\gamma<1$  if the fitted $\beta$ parameter to the X-ray surface 
brightness of the cluster is in the range of from $1/3$ 
to $2/3$. This requires an increasing temperature with cluster radius,
in conflict with the naive speculation that the overall temperature 
profile of a cluster should drop with radius. 
The latter occurs, according to our prediction, 
only inside cluster with $2/3<\beta<1$.  
At present, the spectroscopic analyses have resulted in 
a discrepancy in the radial temperature gradients due primarily
to the problem with the sensitivities and resolutions of detectors. 
For instance, by contrast to the
remarkably radial temperature decline claimed by Markevitch et al. (1998),
the nearly flat temperature profiles have recently been reported by
Irwin, Bregman \& Evrard (1999) based on a detailed comparison of
the ASCA and ROSAT PSPC determined temperature profiles of 26
clusters. In particular, the trend of an increasing temperature 
with radius is likely to present in a few cases 
(e.g. 2A0335, A401, A1651, A3571, etc.).  
Therefore, it is not impossible that the temperature profiles 
would have a mild increase with radii for  clusters with  
$1/3<\beta<2/3$.  A conclusive test for this claim will be  
provided by the space experiments like AXAF and XMM.

Another concern is the poorly determined $\beta$ and $r_c$ 
parameters used in the computations of the baryon
fraction limits $f_{b,lens}$ and $f_{b,xray}$, because $f_b(\infty)$
depends sensitively on these two parameters.  To date, the X-ray surface 
brightness profiles of clusters have not been well measured 
at large radii ($\ga1$ Mpc). Therefore, there may exist 
rather large uncertainties 
in the present fitting of $\beta$ models, apart from the fact that 
different observations at different energy bands might give 
very different values of $\beta$ and $r_c$ even for the same cluster. 
In particular, the $\beta$ model fitting is affected by the
procedure of whether or not the cooling flow regions are reasonably 
excluded. For instance,  from a detailed study of the surface brightness
distributions of 25 cooling flow clusters,  
Vikhlinin, Forman \& Jones (1999) have  shown 
that the $\beta$ values are systematically higher outside 
the cooling flow regions. Alternatively, 
a double $\beta$ model has been often used in recent years 
to represent quantitatively the excess emission 
in the central cores of  cooling flow clusters (e.g. Ikebe et al. 1996;  
Xu et al. 1998; Mohr, Mathiesen \& Evrard 1999),  
which also results in a larger value of $\beta$ for the extended gas 
component. So, it is possible to release the requirement that  
X-ray temperature should increase with radius in order for $f_b(\infty)$ 
to remain constant if large $\beta$ parameter is adopted. 
Furthermore, because our attempt to derive the universal value $f_b(\infty)$
from the central baryon fractions $f_b(0)$ 
depends sensitively on the central gas distributions,
our current results are subject to the presence of the sharp peaks
in emission concentrated in the cores of some clusters.

Nevertheless, if clusters are representative of the Universe, and 
if the entire intracluster gas is in hydrostatic 
equilibrium with the underlying 
gravitational potential of clusters and follows the $\beta$ model, we would 
expect that the X-ray surface brightness and temperature profiles of clusters 
should satisfy the necessary condition eq.(9), which reads  
$\gamma\approx2(1-1/3\beta)$ for $1/3<\beta<1$.
This will be directly testable when these two profiles are well measured 
(e.g. by AXAF, XMM, etc.).
Even if the temperature profile of a cluster is not available, we will still 
be able to obtain its baryon fraction according to eq.(11) or (17) 
by accurately measuring its X-ray surface brightness 
distribution and its central temperature $T_0$, especially for a cluster
with smaller $\beta$ parameter. Applications to  existing
data and future observations will be made in our subsequent work.

%\vskip 1cm

\section*{Acknowledgments}

We gratefully acknowledge the valuable comments and suggestions
by an anonymous referee. This work was supported by 
the National Science Foundation of China, under Grant No. 1972531.

\newpage

\begin{figure*}
\centerline{\hspace{3cm}\psfig{figure=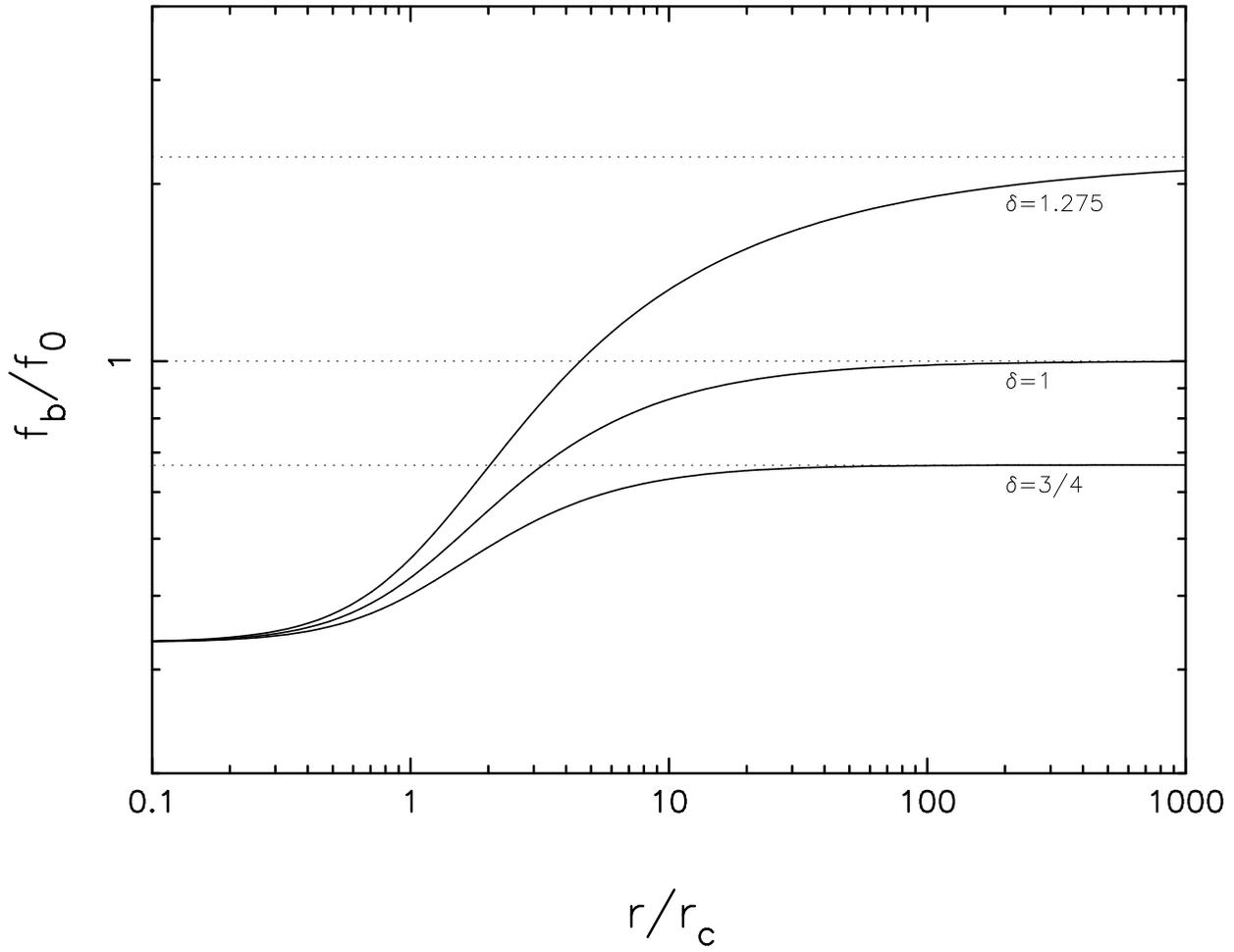,width=1.\textwidth,angle=270}}
\caption{Variations of baryon fractions with cluster radii for
three kinds of gas distributions: $(\delta,\beta)=(3/4,1/2)$, 
$(1,2/3)$ and $(1.275,0.85)$. The corresponding asymptotic values
at large radii are shown by dotted lines. 
} 
\end{figure*}  

\begin{figure*}
\centerline{\hspace{3cm}\psfig{figure=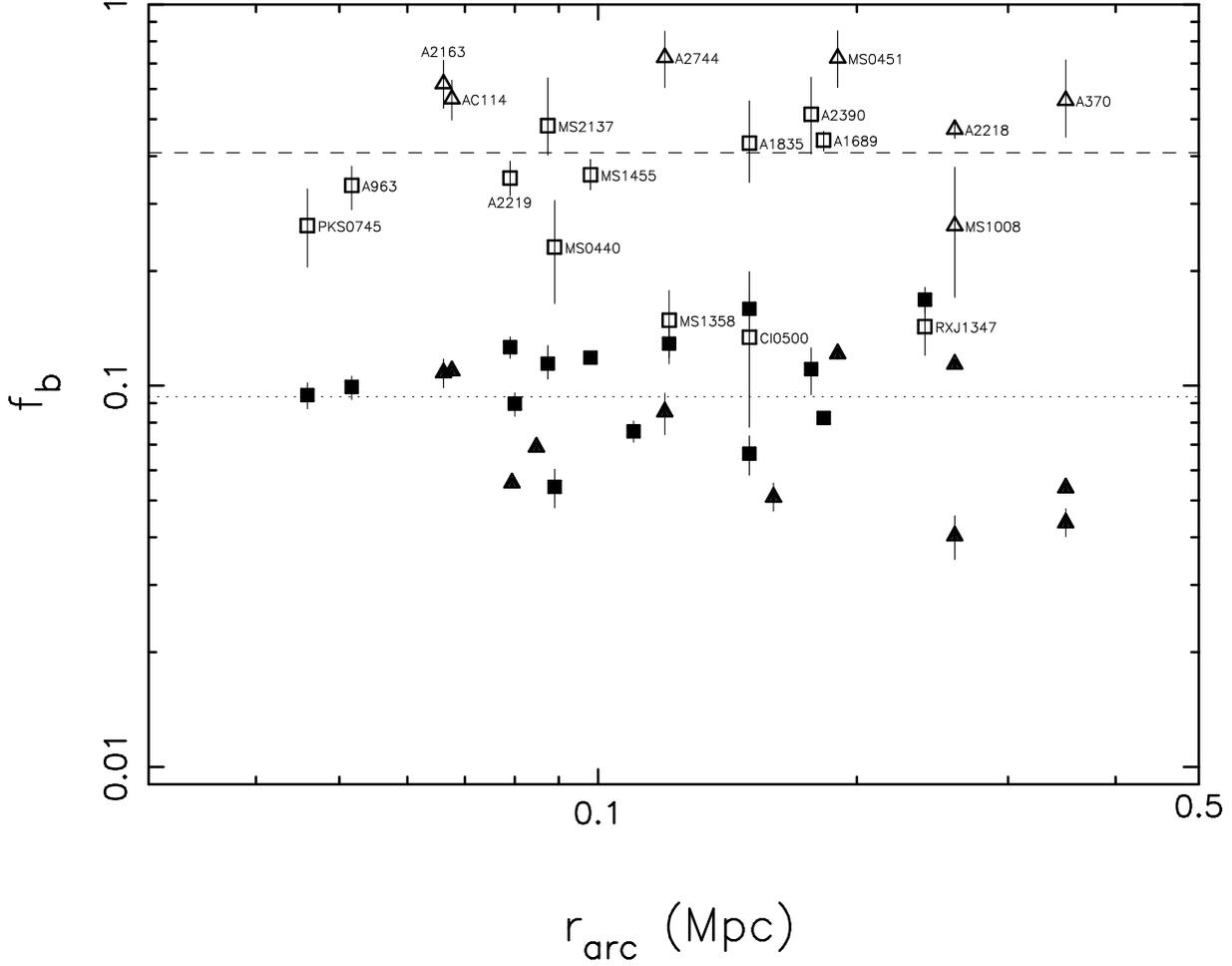,width=1.0\textwidth,angle=270}}
\caption{Asymptotic baryon fractions of 19 
lensing clusters in Table 1 plotted against arc positions $r_{arc}$.
Filled symbols are the low limits derived from the projected baryon
fractions within the 21 arc positions, and open symbols are the
X-ray results derived from $f_b(0)$. For the latter, 
$r_{arc}$ are used to denote the clusters only (we take the first arc 
position for the multiple-arc system). We use the squares to denote
cooling clusters and  the triangles, non-cooling flow ones.
Dotted and dashed lines represent the average values of the lensing and
X-ray results, respectively: $\langle f_{b,lens}\rangle =0.094\pm0.035$ and
 $\langle f_{b,xray}\rangle =0.41\pm0.18$.
}
\end{figure*}

\begin{figure*}
\centerline{\hspace{3cm}\psfig{figure=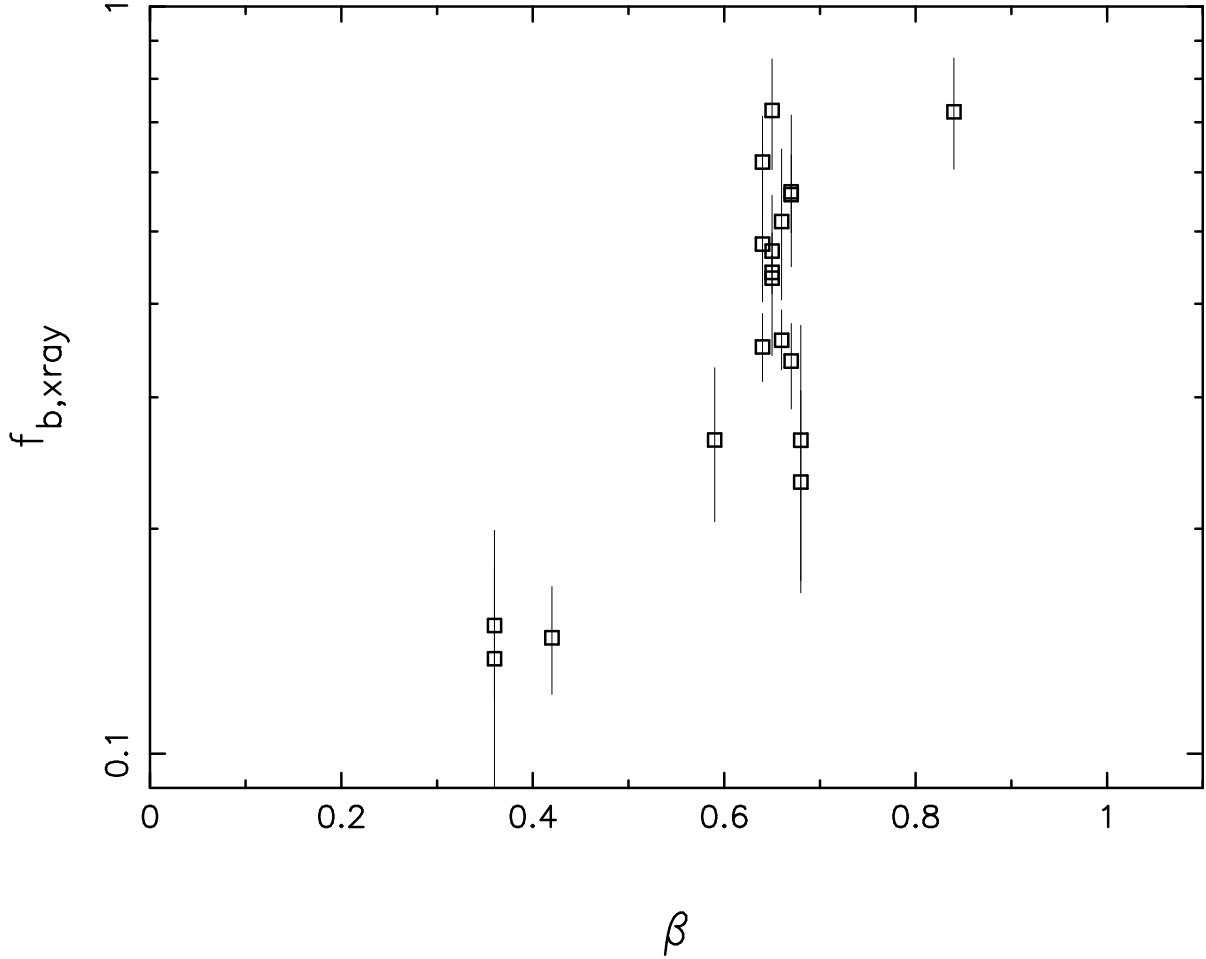,width=1.\textwidth,angle=270}}
\caption{Dependence of asymptotic baryon fractions $f_{b,xray}$
upon $\beta$ parameters for 19 lensing clusters.}
\end{figure*}

\begin{figure*}
\centerline{\hspace{3cm}\psfig{figure=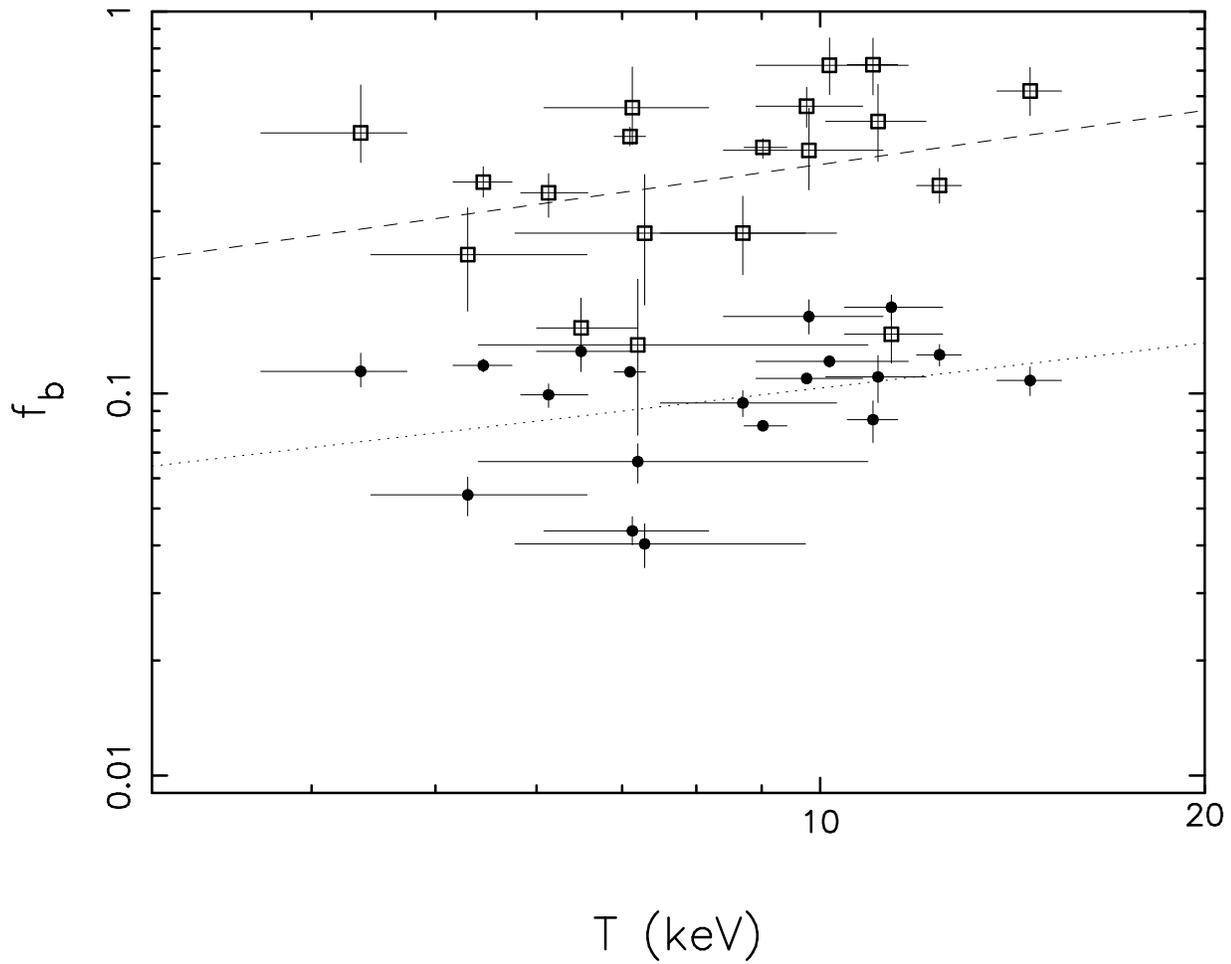,width=1.\textwidth,angle=270}}
\caption{Asymptotic baryon fractions of clusters 
vs. X-ray temperature for 19 lensing clusters. 
Filled circles are the low limits from gravitational lensing, and 
open squares are the X-ray results.
Dotted and dashed lines are the best-fitted 
$f_b-T$ relations to the two sets of data, respectively.} 
\end{figure*}

\end{document}